\documentclass[aps,twocolumn,showpacs]{revtex4}

\newcommand{\stac}[2]{\stackrel{\scriptscriptstyle {#1}}{#2}}

\begin{document}

\title{Cosmological constant and gravitational theory on D-brane}

\author{Tetsuya Shiromizu$^{(1,3)}$, Kazuya Koyama$^{(2)}$ and Takashi Torii$^{(3)}$}

\affiliation{$^{(1)}$Department of Physics, Tokyo Institute of Technology, 
Tokyo 152-8551, Japan}

\affiliation{$^{(2)}$Department of Physics, The University of Tokyo,  Tokyo
113-0033, Japan}

\affiliation{$^{(3)}$Advanced Research Institute for Science and Engineering, 
Waseda University, Tokyo 169-8555, Japan}

\date{\today}

\begin{abstract}
In a toy model we derive the gravitational equation on a self-gravitating 
curved D-brane. 
The effective theory on the brane is drastically changed from the ordinal 
Einstein equation. The net cosmological constant on the brane depends on a 
tuning between the brane tension and the brane charges. Moreover, non-zero matter 
stress tensor exists if the net cosmological constant is not zero. This 
fact indicates a direct connection between  matters on the brane and the 
dark energy. 
\end{abstract}


\maketitle

\label{sec:intro}
\section{Introduction}

Recent observations tell us the existence of the cosmological constant 
or/and dark energy \cite{SN,SN2,MAP}, and  its origin is definitely
important now (See Ref.~\cite{Quitreview} for the review). 
Surprisingly its energy density is almost the same order of magnitude as 
the present mass density of the universe. This fact indicates a connection 
between the dark energy and matter.  In this paper we will focus on it and
show the  direct connection between the presence of the
matter and dark energy in the  brane world context. 

Our idea comes from the brane world based on D-brane. In Ref.~\cite{SKOT}, we
have carefully  investigated the gravitational theory on the {\it
self-gravitating} D-brane in  type IIB supergravity on $S^5$ which may be a
realistic model for D-branes. The  conclusion obtained there is 
that matters do not appear as a source term  of the Einstein equation on the
D-brane. In that analysis we supposed that  the vacuum energy on the brane is
zero. From that observation we can obtain the following  naive conjecture: 
{\it matters are localized on  the brane in gravitational point of view if
there is non-zero vacuum energy on the brane}. That is,  the presence of
matter accompanied by the appearance of the vacuum energy.  
This might be regarded as an evidence of the tight connection between the
matter  and dark energy. 

In the original model of type IIB supergravity on $S^5$, there are undesirable
behavior of  the bulk geometry due to the presence of the scalar fields
related to the  dilaton and the compactification from ten to five dimensions.
Indeed the  single brane model proposed by Randall and Sundrum~\cite{RS}
cannot be realized in the sense that  the bulk geometry is not  warped
enough to make the volume of extra dimensions  finite and the four
dimensional Einstein gravity cannot be recovered even  at low energy. 
Hence
we will discuss in a toy model which is slightly simplified model 
keeping the essence for 
the above remarkable result. 
In our model, one can see that the bulk geometry looks like 
anti-de Sitter spacetime which is appropriate for the Randall-Sundrum II
model. 

The remainder of this paper is organised as follows. In Sec.~\ref{sec:model},
we describe our toy  model and write down the basic equations. In Sec.
\ref{sec:approximation}, we solve the bulk geometry  with the appropriate
boundary conditions on the brane in long-wave approximation \cite{GE}.  Using
the result obtained in Sec.~\ref{sec:model} and \ref{sec:approximation}, we
derive the gravitational theory on the  brane in
Sec.~\ref{sec:D-brane}. The theory has quite different form from the
ordinary theory in four dimensions.  Finally we summarize the present work and
give discussion in Sec.~\ref{sec:summary}.

\section{Model}
\label{sec:model}

\subsection{The action for toy model}

We are interested in the gravitational theory on D3-branes. For simplicity we
work with the toy model 
%
\begin{eqnarray}
S & = & \frac{1}{2\kappa^2} \int d^5x {\sqrt {-g}}\biggl[{}^{(5)}R-2\Lambda
-\frac{1}{2}|H|^2
\nonumber \\
& & -\frac{1}{2}(\nabla \chi)^2-\frac{1}{2}|\tilde F|^2-\frac{1}{2}|\tilde
G|^2 \biggr] +S_{\rm brane}+S_{\rm CS},
\label{action}
\end{eqnarray}
%
where $H_{MNK}=\frac{1}{2}\partial_{[M}B_{NK]}$, 
$F_{MNK}=\frac{1}{2}\partial_{[M}C_{NK]}$, 
$G_{K_1 K_2 K_3 K_4 K_5}=\frac{1}{4!}\partial_{[K_1}D_{K_2 K_3 K_4 K_5]}$, 
$\tilde F = F + \chi H$ and $\tilde G=G+C \wedge H$. $M,N,K=0,1,2,3,4$. 
$B_{MN}$ and $C_{MN}$ are 2-form fields, and $D_{K_1 K_2 K_3 K_4}$ is the
4-form field. 
$S_{\rm brane}$ is given by the Born-Infeld action\footnote{In the present study, 
the terms up to the order of $O({\cal F}^2)$ are enough. The higher order correction 
will be not significant in our current approximation.}  
%
\begin{eqnarray}
S_{\rm brane}=\beta \int d^4x {\sqrt {-{\rm det}(h+{\cal F})}},
\end{eqnarray}
%
where $h_{\mu\nu}$ is the induced metric on the D-brane and 
%
\begin{eqnarray}
{\cal F}_{\mu\nu}=B_{\mu\nu}+(-\beta)^{-1/2}F_{\mu\nu},
\end{eqnarray}
%
and $F_{\mu\nu}$ is the $U(1)$ gauge field on the brane. $\mu,\nu=0,1,2,3$. 
$S_{\rm CS}$ is Chern-Simon action 
%
\begin{eqnarray}
S_{\rm CS} & = & \gamma \int d^4x {\sqrt {-h}}
\epsilon^{\mu\nu\rho\sigma}\biggl[ \frac{1}{4}{\cal
F}_{\mu\nu}C_{\rho\sigma}+\frac{\chi}{8}{\cal F}_{\mu\nu}{\cal
F}_{\rho\sigma}
\nonumber \\
& & +\frac{1}{24}D_{\mu\nu\rho\sigma} \biggr].
\end{eqnarray}
%
In general $\gamma \neq \beta$. We will show the difference is essential to
recover the  ordinary four dimensional Einstein gravity on the brane. 
Setting 
%
\begin{eqnarray}
2\Lambda +\frac{5\kappa^4}{6}\gamma^2=0
\end{eqnarray}
%
as well as $\beta=\gamma$,
we obtain the results  in Ref.~\cite{SKOT}.

In our toy model (\ref{action}), there are no scalar fields 
related to the dilaton and compactification while the model in Ref.
\cite{SKOT} includes these fields.
This is because we can 
see that the scalar fields do not play an important role 
but make the calclations complicated in Ref. 
\cite{SKOT}. 
Instead we introduced 
the bulk cosmological constant $\Lambda$. 
On the other hand, the form fields are indispensable.
As seen soon the background solution could be anti-de Sitter like  
spacetime due to the current simplification of the action and 
it is guaranteed that 
the four dimensional Einstein gravity is recovered on the brane. 

For simplicity we set $H_{\mu\nu\alpha}=0$ and $\tilde F_{\mu\nu\alpha}=0$.

\subsection{Basic equations}

In this subsection we write down the basic equations and boundary conditions. 
Since  we are interested in the effective theory on the brane, it is better
for our purpose to adopt the bulk metric 
%
\begin{eqnarray}
ds^2=dy^2+g_{\mu\nu}(y,x) dx^\mu dx^\nu,
\end{eqnarray}
%
and  perform (1+4)-decomposition. $y$ is the coordinate orthogonal to the 
brane. 
 
The ``evolutional" equations to the $y$-direction are 
%
\begin{eqnarray}
\partial_y K 
= R-\kappa^2 \biggl( {}^{(5)}T^\mu_\mu -\frac{4}{3}{}^{(5)}T^M_M \biggr) -K^2,
\end{eqnarray}
%
%
\begin{eqnarray}
\partial_y \tilde K^\mu_\nu = \tilde R^\mu_\nu -\kappa^2\biggl({}^{(5)}T^\mu_\nu
-\frac{1}{4}
\delta^\mu_\nu {}^{(5)}T^\alpha_\alpha \biggr)-K \tilde K^\mu_\nu,
\label{traceless}
\end{eqnarray}
%
%
\begin{eqnarray}
\partial_y^2 \chi +D^2 \chi +K\partial_y \chi-\frac{1}{2}H_{y\alpha\beta}\tilde 
F^{y\alpha\beta}=0,
\end{eqnarray}
%
%
\begin{eqnarray}
\partial_y X^{y\mu\nu}+KX^{y\mu\nu}
+\frac{1}{2}F_{y\alpha\beta}\tilde G^{y\alpha\beta\mu\nu}=0,
\end{eqnarray}
%
%
\begin{eqnarray}
\partial_y \tilde F^{y\mu\nu}+K \tilde F^{y\mu\nu}-\frac{1}{2}H_{y\alpha\beta}
\tilde G^{y\alpha\beta\mu\nu}=0,
\end{eqnarray}
%
%
\begin{eqnarray}
\partial_y \tilde G_{y \alpha_1 \alpha_2 \alpha_3 \alpha_4}
=K\tilde  G_{y \alpha_1 \alpha_2 \alpha_3 \alpha_4},
\end{eqnarray}
%
where $X^{y\mu\nu}:=H^{y\mu\nu}+\chi \tilde F^{y\mu\nu}$ and the 
energy-momentum tensor is
%
\begin{eqnarray}
&& \kappa^2\;{}^{(5)\!}T_{MN} =  \frac{1}{2}\biggl[ \nabla_M \chi \nabla_N \chi
-\frac{1}{2}g_{MN} (\nabla \chi)^2 \biggr]
\nonumber \\
& & ~~~~~~~~~~
+\frac{1}{4}\biggl[H_{MKL}H_N^{~KL}-g_{MN}|H|^2 \biggr] 
\nonumber \\
& & ~~~~~~~~~~
 +\frac{1}{4}\biggl[\tilde F_{MKL}\tilde
F_N^{~KL}-g_{MN}|\tilde F|^2
\biggr]
\nonumber \\
& & ~~~~~~~~~~
 +\frac{1}{96}\tilde G_{MK_1 K_2 K_3 K_4} \tilde G_{N}^{~~K_1
K_2 K_3 K_4}-\Lambda g_{MN}.
\nonumber \\
& & 
\end{eqnarray}
%
$K_{\mu\nu}$ is the extrinsic curvature, $K_{\mu\nu}=\frac{1}{2}\partial_y g_{\mu\nu}$. 
$\tilde K^\mu_\nu$ and $\tilde R^\mu_\nu$ are the traceless parts
of $K^\mu_\nu$ and $R^\mu_\nu$, respectively.

The constrains are 
%
\begin{eqnarray}
-\frac{1}{2}\biggl[R-\frac{3}{4}K^2+\tilde K^\mu_\nu \tilde K^\nu_\mu \biggr]
=\kappa^2\:{}^{(5)\!}T_{yy},
\end{eqnarray}
%
%
\begin{eqnarray}
D_\nu K^\nu_\mu-D_\mu K = \kappa^2\:{}^{(5)\!}T_{\mu y},
\end{eqnarray}
%
%
\begin{eqnarray}
D^\alpha X_{y\alpha\mu}=0,
\end{eqnarray}
%
%
\begin{eqnarray}
D^\alpha F_{y\alpha\mu}=0,
\end{eqnarray}
%
%
\begin{eqnarray}
D^\alpha \tilde G_{y \alpha \mu_1 \mu_2 \mu_3}=0,
\end{eqnarray}
%
where $D_\mu$ is the covariant derivative with respect to $g_{\mu\nu}$.

The junction conditions at the brane located $y=0$ are 
%
\begin{eqnarray}
\Bigl[K_{\mu\nu}-h_{\mu\nu}K\Bigr]_{y=0} & = & 
-\frac{\kappa^2}{2}\beta
(h_{\mu\nu}-T_{\mu\nu}) +O(T_{\mu\nu}^2)
\nonumber \\
& = & 
-\frac{\kappa^2}{2}\gamma
(h_{\mu\nu}-T_{\mu\nu}) 
\nonumber \\ & & -\frac{\kappa^2}{2}(\beta-\gamma)
(h_{\mu\nu}-T_{\mu\nu})+O(T_{\mu\nu}^2),
 \nonumber \\ & & 
\end{eqnarray}
%
%
\begin{eqnarray}
H_{y\mu\nu}(0,x)=-\kappa^2 \beta {\cal F}_{\mu\nu},
\end{eqnarray}
%
%
\begin{eqnarray}
\tilde F_{y\mu\nu}(0,x)
=-\frac{\kappa^2}{2}\gamma \epsilon_{\mu\nu\alpha\beta}{\cal F}^{\alpha\beta},
\end{eqnarray}
%
%
\begin{eqnarray}
\tilde G_{y\mu\nu\alpha\beta}(0,x)
=- \kappa^2 \gamma \epsilon_{\mu\nu\alpha\beta},
\end{eqnarray}
%
%
\begin{eqnarray}
\partial_y \chi (0,x) 
= -\frac{\kappa^2}{8}\gamma \epsilon^{\mu\nu\alpha\beta}{\cal F}_{\mu\nu}{\cal
F}_{\alpha\beta},
\end{eqnarray}
%
where 
%
\begin{eqnarray}
T_{\mu\nu}={\cal F}_{\mu\alpha}{\cal F}^{~~\alpha}_\nu -\frac{1}{4}h_{\mu\nu}
{\cal F}_{\alpha\beta} {\cal F}^{\alpha\beta}.
\end{eqnarray}
%
The boundary conditions are essential for self-gravitating brane.

\section{Long-wave approximation}
\label{sec:approximation}

To derive the effective gravitational theory on the brane, 
the effects from the bulk are also important and it is necessary
to solve the bulk fields. 
In this paper we employ the gradient 
expansion scheme as Ref.~\cite{SKOT}. 

We assume the following hierarchy in the order of magnitude 
%
\begin{eqnarray}
|T_{\mu\nu}^2| < |T_{\mu\nu}(1-\beta/\gamma)|  
< (1-\beta/\gamma)^2 \ll |T_{\mu\nu}| < |1-\beta/\gamma| . 
\end{eqnarray}
%
Note that we assume $\beta\leq\gamma<0$.
We will derive the gravitational theory up to the order of 
$O(T_{\mu\nu}(\gamma-\beta))$, that is, we will omit $O(T_{\mu\nu}^2)$ terms 
for simplicity. As we will see soon, $R_{\mu\nu}(h) =O(\gamma-\beta)$. 

The bulk metric is written in Gaussian-normal coordinate again
%
\begin{eqnarray}
ds^2=dy^2+g_{\mu\nu}(y,x) dx^\mu dx^\nu. 
\end{eqnarray}
%
The induced metric on the brane will be denoted by 
$h_{\mu\nu}:=g_{\mu\nu}(0,x)$ 
and then
%
\begin{eqnarray}
g_{\mu\nu}(y,x)
=a^2(y)\Bigl[h_{\mu\nu}(x)+\stac{(1)}{g}_{\mu\nu}(y,x)+\cdots\Bigr].
\end{eqnarray}
%
In the above $\stac{(1)}{g}_{\mu\nu}(0,x)=0 $ and $a(0)=1$. In a similar way, 
the extrinsic curvature is expanded as 
%
\begin{eqnarray}
K^\mu_\nu = \stac{(0)}{K^\mu_\nu}+ \stac{(1)}{K^\mu_\nu}+\stac{(2)}{K^\mu_\nu}+
\cdots.
\end{eqnarray}
%

\subsection{0th order}

It is easy to obtain the zeroth order solutions. Without derivation we present
them.
%
\begin{eqnarray}
\stac{(0)}{K^\mu_\nu}=-\frac{1}{\ell}\delta^\mu_\nu,
\end{eqnarray}
%
%
\begin{eqnarray}
\stac{(0)}{R}_{\mu\nu}(h)=0,
\end{eqnarray}
%
%
\begin{eqnarray}
\stac{(0)}{g}_{\mu\nu}=a^2(y)h_{\mu\nu}(x)=e^{-\frac{2}{\ell}y}h_{\mu\nu}(x),
\end{eqnarray}
%
where 
%
\begin{eqnarray}
\frac{1}{\ell}=-\frac{1}{6}\kappa^2 \gamma.
\end{eqnarray}
%
$\ell$ is the curvature scale of anti-deSitter like spacetimes. 
This represents the Randall-Sundrum tuning. 

In addition,
%
\begin{eqnarray}
\tilde G_{y\alpha_1 \alpha_2 \alpha_3 \alpha_4}=-a^4 \kappa^2 
\gamma \epsilon_{\alpha_1 \alpha_2 \alpha_3 \alpha_4},
\end{eqnarray}
%
where $\epsilon_{\alpha_1 \alpha_2 \alpha_3 \alpha_4}$ is the Levi-Civita 
tensor with respect to the induced metric $h_{\mu\nu}$ on the 
brane\footnote{$G_{y\alpha_1\alpha_2\alpha_3\alpha_4}$ can be solved in full order:
$G_{y\alpha_1\alpha_2\alpha_3\alpha_4}=-\kappa^2 \gamma 
\epsilon_{\alpha_1\alpha_2\alpha_3\alpha_4}(g)$. 
Here $\epsilon_{\alpha_1 \alpha_2 \alpha_3 \alpha_4}(g)$ is the Levi-Civita 
tensor with respect to $g_{\mu\nu}$.}. 
The warp factor $a(y)$ behaves well for the localization of 
gravity on the brane, that is, we do not encounter a serious 
problem of the localization in our previous work~\cite{SKOT}.

\subsection{1st order}

The first order equations for $\tilde F_{y\mu\nu}$ and $H_{y\mu\nu}$ are 
%
\begin{eqnarray}
\partial_y \stac{(1)}{\tilde F}_{y\mu\nu}-\frac{1}{2a^4}
\stac{(1)}{H}_{y\alpha\beta} \tilde G_{y\rho\sigma\mu\nu}h^{\alpha\rho}h^{\beta\sigma}=0
\end{eqnarray}
%
and
%
\begin{eqnarray}
\partial_y \stac{(1)}{H}_{y\mu\nu}+\frac{1}{2a^4}\stac{(1)}{\tilde F}_{y\alpha\beta}\tilde 
G_{y\rho\sigma\mu\nu}h^{\alpha\rho}h^{\beta\sigma}=0.
\end{eqnarray}
%
Together with the junction conditions the solutions are given by 
%
\begin{eqnarray}
\stac{(1)}{H}_{y\mu\nu}=-\kappa^2 \gamma a^{-6} {\cal F}_{\mu\nu}
\end{eqnarray}
%
and
%
\begin{eqnarray}
\stac{(1)}{\tilde F}_{y\mu\nu}=-\frac{\kappa^2}{2}\gamma a^{-6} 
\epsilon_{\mu\nu\rho\sigma}{\cal F}_{\alpha\beta}h^{\rho\alpha}h^{\sigma\beta}.
\end{eqnarray}
%
Using these results the evolutional equation for the traceless part 
of the extrinsic curvature is 
%
\begin{eqnarray}
\partial_y \stac{(1)}{\tilde K^\mu_\nu}=
-\stac{(0)}{K}\stac{(1)}{\tilde K^\mu_\nu}+\frac{1}{a^2}
\tilde R^\mu_\nu (h) -\kappa^4 \gamma^2 a^{-16} T^\mu_\nu, 
\end{eqnarray}
%
where $R^\mu_\nu (h)=h^{\mu\alpha}R_{\alpha\nu}(h)$ 
is the Ricci tensor with respect to $h_{\mu\nu}$
and $T^\mu_\nu= h^{\mu\alpha}T_{\alpha\nu}$. The solution is 
%
\begin{eqnarray}
\stac{(1)}{\tilde K^\mu_\nu}(y,x)=-\frac{\ell}{2a^2}\tilde R^\mu_\nu (h) 
+\frac{1}{2}\kappa^2 \gamma  a^{-16}T^\mu_\nu + \frac{\chi^\mu_\nu (x)}{a^4},
\end{eqnarray}
%
where $\chi^\mu_\nu$ is the ``integration of constant" and expresses 
the holographic CFT stress tensor\cite{holo}. 
Since it is not affect the result below, we will omit $\chi^\mu_\nu$ hereafter. 

The trace part of the extrinsic curvature can be evaluated 
from the Hamiltonian constraint as 
%
\begin{eqnarray}
\stac{(1)}{K}(y,x)=-\frac{\ell}{6a^2} R(h).
\end{eqnarray}
%
Finally, we obtain
%
\begin{eqnarray}
\bigl[K^\mu_\nu -\delta^\mu_\nu K\bigr]^{(1)}=-\frac{\ell}{2a^2}
G^\mu_\nu (h)+\frac{1}{2}\kappa^2 \gamma a^{-16} T^\mu_\nu.
\end{eqnarray}
%
From junction condition, it is easy to see that the Einstein equatoin up to 1st order becomes 
%
\begin{eqnarray}
G_{\mu\nu}(h)=-\frac{\kappa^2}{\ell} (\gamma-\beta)h_{\mu\nu}. 
\end{eqnarray}
%
The stress energy tensor of the gauge field does not appear in this order. Thus we 
must discuss the next order to see how the ordinal gravitational equation recovers. 

We calculate the metric at first order,
$\stac{(1)}{g}_{\mu\nu}$, which will be  used in the computation of the second
order Ricci tensor
$[R^\mu_\nu]^{(2)}$. The result is 
%
\begin{eqnarray}
\stac{(1)}{g}_{\mu\nu} & = & \frac{\ell^2}{2}(1-a^{-2}) 
 \biggl[ R_{\mu\nu}(h)-\frac{1}{6}h_{\mu\nu}R(h) \biggr] \nonumber \\
& & +\frac{3}{8}(1-a^{-16})T_{\mu\nu}.\label{metric}
\end{eqnarray}
%

\subsection{2nd order}

First of all, using Eq.~(\ref{metric}), we compute the second order Ricci tensor 
%
\begin{widetext}
\begin{eqnarray}
\bigl[R^\mu_\nu  \bigr]^{(2)}
& = &  
\frac{\ell^2}{4}a^{-2}(1-a^{-2} )\biggl[ 
D_\alpha D_\nu R^\alpha_\mu +D_\alpha D_\mu R^\alpha_\nu 
-D^2 R^\mu_\nu -\frac{2}{3}D^\mu D_\nu R +\frac{1}{6}\delta^\mu_\nu D^2 R
-2 R^\mu_\alpha R^\alpha_\nu +\frac{1}{3}R^\mu_\nu R 
\biggr] 
\nonumber \\
& & ~~~~~
+\frac{3}{16}a^{-2}(1-a^{-16})
\biggl[ 
D_\alpha D_\nu T^{\alpha\mu} +D_\alpha D^\mu T^\alpha_\nu
-D^2 T^\mu_\nu - 2 T^\mu_\alpha R^\alpha_\nu 
\biggr] 
\nonumber \\
& \simeq &
a^{-2}\biggl[
\frac{\kappa^4}{6}(\gamma-\beta)^2 (a^{-2}-1)\delta^\mu_\nu
+\frac{3\kappa^2}{8\ell}(\gamma-\beta)(a^{-16}-1)T^\mu_\nu
-\frac{3}{16}(a^{-16}-1) \Bigl( D_\alpha D_\nu T^{\alpha\mu} 
+D_\alpha D^\mu T^\alpha_\nu -D^2 T^\mu_\nu  \Bigr) 
\biggr].
\nonumber \\
\end{eqnarray}
\end{widetext}
%
From the first to second lime in the r.h.s we used $R_{\mu\nu}(h) \simeq
\frac{\kappa^2}{\ell}(\gamma-\beta) h_{\mu\nu}$. The second term in
Eq.~(\ref{traceless}) is computed as 
%
\begin{eqnarray}
& & \kappa^2\biggl[{}^{(5)}T^\mu_\nu -\frac{1}{4}\delta^\mu_\nu {}^{(5)}T^\alpha_\alpha 
\biggr]^{(2)} 
\nonumber \\
& & ~~~=-\kappa^4 \gamma (\gamma-\beta)(2a^{-18}-a^{-16})T^\mu_\nu. 
\label{traceless2}
\end{eqnarray}
%
Since we need the higher order solutions for $\tilde F_{y\mu\nu}$ and 
$H_{y\mu\nu}$, we derive them in the appendix. 
Then the evolutional equation for $\stac{(2)}{\tilde K^\mu_\nu}$ is 
%
\begin{eqnarray}
\partial_y \stac{(2)}{\tilde K^\mu_\nu} 
& = &  \Bigl[ \tilde R^\mu_\nu\Bigr]^{(2)}-\stac{(1)}{K}
\stac{(1)}{\tilde K^\mu_\nu}-\stac{(0)}{K}\stac{(2)}{\tilde K^\mu_\nu}
\nonumber \\
& & 
-\kappa^4 \gamma (\beta-\gamma)(2a^{-18}-a^{-16}) T^\mu_\nu.
\end{eqnarray}
%
Its solution is easily found and the value on the brane becomes 
%
\begin{eqnarray}
\stac{(2)}{\tilde K^\mu_\nu} & = & 
 -\frac{3\ell}{28}\biggl[
D_\alpha D_\nu T^\alpha_\mu +D_\alpha D_\mu T^\alpha_\nu
-D^2 T^\mu_\nu \biggr]\nonumber \\
& & +\frac{\ell}{21}\kappa^4\gamma (\gamma -\beta)T^\mu_\nu.
\end{eqnarray}
%
 From the Hamiltonian constraint
%
\begin{eqnarray}
& & -\Bigl[R(h)\Bigr]^{(2)}+\frac{3}{2}
\stac{(0)}{K}\stac{(2)}{K}+\frac{3}{4}\stac{(1)}{K^2}
-\stac{(1)}{\tilde K^\mu_\nu}\stac{(1)}{\tilde K^\nu_\mu}
\nonumber \\ & & ~~~~~~~
=\frac{1}{2}\kappa^4 \gamma (\beta-\gamma){\cal
F}_{\mu\nu}{\cal F}^{\mu\nu},
\end{eqnarray}
%
we can compute the trace part of second order extrinsic curvature
$\stac{(2)}{K^\mu_\nu}$
%
\begin{eqnarray}
\stac{(2)}{K} & = & -\frac{\ell}{2}\tilde R^\mu_\nu 
T^\nu_\mu-\frac{\ell}{12}\kappa^4\gamma(\beta-\gamma){\cal F}_{\mu\nu}{\cal F}^{\mu\nu}
\nonumber \\
& & -\frac{\ell^3}{24}\biggl(R^\mu_\nu R^\nu_\mu -\frac{1}{3}R^2 \biggr) \nonumber \\
& \simeq &  \frac{\kappa^4\ell}{18}(\gamma-\beta)^2+\frac{\kappa^4\ell}{12}
\gamma (\gamma-\beta){\cal F}_{\mu\nu}{\cal F}^{\mu\nu} .
\end{eqnarray}
%

\section{Low energy gravitational theory on D-brane}
\label{sec:D-brane}

\subsection{General case}

Now we are ready to derive the gravitational equation on the brane. 
{}From the junction condition, the effective equation is given by 
%
\begin{eqnarray}
G_{\mu\nu}(h) & = & - \frac{\kappa^2}{\ell}(\gamma-\beta) h_{\mu\nu}
-\frac{\kappa^4}{12}(\gamma-\beta)^2 h_{\mu\nu} \nonumber \\
& & +\frac{3}{7}\frac{\kappa^2}{\ell} \biggl(1-\frac{\gamma}{\beta} \biggr)
\stac{(F)}{T}_{\mu\nu}
\nonumber \\
& & +\frac{3}{4}\frac{\kappa^2}{\ell} \biggl(1-\frac{\gamma}{\beta}\biggr)
F_{\alpha\beta}F^{\alpha\beta}h_{\mu\nu} +\tau_{\mu\nu},
\label{eq-brane}
\end{eqnarray}
%
where we set $B_{\mu\nu}=0$ and defined 
%
\begin{eqnarray}
\stac{(F)}{T}_{\mu\nu}:=F_{\mu\alpha}F^{~\alpha}_\nu -\frac{1}{4}h_{\mu\nu} 
F_{\alpha\beta} 
F^{\alpha\beta}.
\end{eqnarray}
%
$\tau_{\mu\nu}$ is defined by 
%
\begin{eqnarray}
\tau_{\mu\nu}=\frac{3}{14\beta}\biggl(D_\alpha D_\nu \stac{(F)}{T^\alpha_\mu} 
+D_\alpha D_\mu \stac{(F)}{T^\alpha_\nu} -D^2 \stac{(F)}{T}_{\mu\nu}\biggr).
\end{eqnarray}
%
Note that $\tau_{\mu\nu}$ is traceless. 

The first and second terms in the r.h.s of Eq.~(\ref{eq-brane}) are regarded
as the (positive) vaccum energy. The third term is the energy-momentun tensor of the
gauge field with the coupling which depends  on $\beta$,
$\gamma$ and $\ell$. 

When $\gamma=\beta$, the equation becomes vacuum Einstein equation
%
\begin{eqnarray}
G_{\mu\nu}(h) \simeq 0.
\end{eqnarray}
%
This is the result obtained in our previous study \cite{SKOT}. 
In the above we used $\tau_{\mu\nu} \propto (\gamma-\beta)$.

\subsection{Homogeneous and isotropic universe}

Let us examine the gravitational equation on the brane by considering
the homogeneous and isotropic universe. The gauge field on the brane 
will be regarded as the radiatoin fluid. In this case, $F^2 \simeq 0$ 
and then the gravitational equation becomes
%
\begin{eqnarray}
G_{\mu\nu}(h) \simeq -\lambda_{\rm eff} h_{\mu\nu} + S_{\mu\nu}^{\rm eff},
\end{eqnarray}
%
where 
%
\begin{eqnarray}
\lambda_{\rm eff}:= \frac{\kappa^2}{\ell}(\gamma-\beta)+
\frac{\kappa^4}{12}(\gamma-\beta)^2 ,
\end{eqnarray}
%
and
%
\begin{eqnarray}
S_{\mu\nu}^{\rm eff}:=\frac{3}{7}\frac{\kappa^2}{\ell}
\biggl(1-\frac{\gamma}{\beta} \biggr)T_{\mu\nu}+\tau_{\mu\nu} .
\end{eqnarray}
%
Since $S_{\mu\nu}^{\rm eff}$ is  traceless, it behaves like radiation 
in homogeneous and isotropic universe.
%
\begin{eqnarray}
S_{00}^{\rm eff} \simeq 
\frac{1}{7}
\frac{\kappa^2}{\ell}
\biggl(1-\frac{\gamma}{\beta} \biggr)\rho,
\end{eqnarray}
%
where $\rho=\stac{(F)}{T}_{00}$. 

Consequently we can observe that the conventional cosmology is recovered if
and only if the non-zero (positive) vacuum energy exists.

\section{Summary and discussion}
\label{sec:summary}

In this paper, we derive the effective gravitational theory on a self-gravitating 
D-brane in the toy model. The 
brane is described by (the bosonic part of) Born-Infeld action and then we
expect naively that  the gauge field on the brane acts as the source term of
the four dimensional Einstein gravity. The result is as
follows. If the net vacuum energy is tuned to zero on the brane,  the gauge
field does not act as the source term in the gravitational
theory  on the brane. On the other hand, if the net positive vacuum energy is
on the brane,  the gauge field plays the source term on the brane. 
This implies that the existence of the cosmological constant on the brane is 
essential for the realization of the four dimensional Einstein gravity on the 
brane. This might give a new insight into a direct connection between matters
and dark energy. 

Finally we briefly comment on the unfamiliar term $h_{\mu\nu}F^2$ in Eq.
(\ref{eq-brane}). 
It is cosmology where gauge fields affect the spacetime in
observational point of view.  
Under the fluid approximation in the
homogeneous and isotropic universe, however, 
$F^2 \simeq 0$. Thus the existence of such unfamilar term is not incosistent with 
observations. To judge our current model the study on fermionic part is desired. 

\section*{Acknowledgments}

We would like to thank Nathalie Deruelle, Kenji Hotta, Yousuke Imamura, Katsushi Ito, 
Kei-ichi Maeda, Sumitada Onda and Misao Sasaki for fruitful discussions. To complete this work, the
discussion during and after the YITP workshops YITP-W-01-15 and  YITP-W-02-19
were useful. The work of TS was supported by Grant-in-Aid for Scientific
Research from Ministry of Education, Science, Sports and Culture of 
Japan(No.13135208, No.14740155 and No.14102004). The work of KK was supported by JSPS.

\vskip 1cm 

\appendix

\section{Derivation of Eq. (\ref{traceless2})}

In the next order the equations for $\tilde F_{y\mu\nu}$ and $H_{y\mu\nu}$ becomes 
%
\begin{eqnarray}
\partial_y \stac{(2)}{\tilde F}_{y\mu\nu}-\frac{3}{\ell} \stac{(2)}{H}_{y\alpha\beta}
\epsilon^{\alpha\beta}_{~~~\mu\nu}=0
\end{eqnarray}
%
and
%
\begin{eqnarray}
\partial_y \stac{(2)}{\tilde H}_{y\mu\nu}+\frac{3}{\ell} \stac{(2)}{\tilde 
F}_{y\alpha\beta}
\epsilon^{\alpha\beta}_{~~~\mu\nu}=0.
\end{eqnarray}
%
The junction conditions are 
%
\begin{eqnarray}
\stac{(2)}{\tilde F}_{y\alpha\beta}(0,x)=0
\end{eqnarray}
%
and
%
\begin{eqnarray}
\stac{(2)}{H}_{y\alpha\beta}(0,x)=\kappa^2 (\gamma-\beta){\cal F}_{\mu\nu}.
\end{eqnarray}
%
It is easy to obtain the solutions as 
%
\begin{eqnarray}
\stac{(2)}{\tilde F}_{y\alpha\beta}=\frac{\kappa^2}{4}(\gamma-\beta)(a^{-6}-a^6)
\epsilon_{\mu\nu\alpha\beta}{\cal F}^{\alpha \beta}
\end{eqnarray}
%
and
%
\begin{eqnarray}
\stac{(2)}{H}_{y\alpha\beta}=\frac{\kappa^2}{2}(\gamma-\beta)(a^{-6}+a^6){\cal F}_{\mu 
\nu}.
\end{eqnarray}
%
Then we obtain the expression of Eq. (\ref{traceless2}).


\begin{thebibliography}{22}

\bibitem{SN}
S. Perlmutter {\it et al.}, Astrophys. J. {\bf 483}, 565 (1997);
S. Perlmutter {\it et al.}, Astrophys. J. {\bf 517}, 565 (1999).

\bibitem{SN2}
A. Riess {\it et al.}, Astron. J. {\bf 116}, 1009 (1998).

\bibitem{MAP}
C. L. Bennett {\it et. al.}, astro-ph/0302207.

\bibitem{Quitreview}
For the review, P. J. E. Peebles and B. Ratra, Rev. Mod. Phys. {\bf 75}, 559(2003);
T. Padmanabhan, hep-th/0212290(to appear in Phys. Rept.).

\bibitem{SKOT}
T. Shiromizu, K. Koyama, S. Onda and T. Torii, hep-th/0305253, to appear in Phys. Rev. {\bf D}(2003).

\bibitem{RS}
L.~Randall and R.~Sundrum, Phys. Rev. Lett. {\bf 83}, 4690 (1999).

\bibitem{GE}
T. Wiseman, Class. Quant. Grav. {\bf 19}, 3083(2002);
S. Kanno and J. Soda, Phys. Rev. D{\bf 66}, 043526(2002);
{\rm ibid}, 083506,(2002);
T. Shiromizu and K. Koyama, Phys. Rev. D{\bf 67}, 084022(2003);
S. Kanno and J. Soda, hep-th/0303203. 

\bibitem{holo}
S. S. Gubser, Phys. Rev. {\bf D63}, 084017(2001);
L. Anchordoqui, C. Nunez and L. Olsen, JHEP {\bf 10}, 050(2000);
S.~B.~Giddings, E.~Katz, and L.~Randall, JHEP {\bf 0003}, 023(2000);
T.~Shiromizu and D.~Ida, Phys. Rev. {\bf D64}, 044015(2001);
S. de Haro, K. Skenderis, and S. N. Solodukhin, hep-th/0011230;
T.~Shiromizu, T.~Torii, and D.~Ida, JHEP {\bf 0203},007(2002);
S. Nojiri, S.D. Odintsov, S. Zerbini, Phys. Rev. {\bf D62},064006(2000);
S. Nojiri, S.D. Odintsov, Phys. Lett. {\bf B484}, 119(2000);
S. W. Hawking, T. Hertog and H. S. Reall, Phys. Rev. D{\bf 62}, 043501 (2000);
K. Koyama and J. Soda, JHEP {\bf 05}, 027 (2001).

\end{thebibliography}
\end{document}